\documentclass{article}

\usepackage{amsfonts,amssymb, amsmath,mathrsfs}
\usepackage{graphicx}

\usepackage[english]{babel}

\DeclareMathAlphabet{\mathpzc}{OT1}{pzc}{m}{it}

\def\on#1#2{\mathop{\vbox{\ialign{##\crcr\noalign{\kern2pt}
$\scriptstyle{#2}$\crcr\noalign{\kern2pt\nointerlineskip}
\kern-2pt$\hfil\displaystyle{#1}\hfil$\crcr}}}\limits}

\def\nn{ \nonumber }
\def\bq{ \begin{equation} }
\def\eq{ \end{equation} }
\def\ben{ \begin{eqnarray} }
\def\en{ \end{eqnarray} }

\def\e{{\rm e}}

\newtheorem{prop}{Proposition}
\newtheorem{exa}{Example}
\newenvironment{exam}{\begin{exa} \rm }{\end{exa}}

\newtheorem{cas}{Case}
\newenvironment{case}{\begin{cas} \rm }{\end{cas}}

\newtheorem{re}{Remark}
\newenvironment{rem}{\begin{re} \rm }{\end{re}}

\begin{document}

%%%%%%%%%%%% TITLE %%%%%%%%%%%%%%

\title{Leonard Euler: addition theorems and superintegrable systems}

\author{A V Tsiganov \\
\it\small
St.Petersburg State University, St.Petersburg, Russia\\
\it\small e--mail: tsiganov@mph.phys.spbu.ru}
\date{}
 \maketitle

\begin{abstract}
 We consider the Euler approach to construction and to investigation of the superintegrable systems related to the addition theorems. As an example we reconstruct Drach systems and get some new two-dimensional superintegrable St\"ackel systems.
\end{abstract}

\section{Introduction}
There are many different mathematical approaches to classification and to investigation of the superintegrable systems  with very extensive literature on the subject, see some recent rewiews \cite{cd06,mw08,kkm07}. Here we want to discuss
one of the oldest but almost completely forgotten approaches, which  relates superintegrable systems with the addition theorems.

The story began in 1761  when  Euler proposed a generic construction of the additional algebraic integral of equations of motion
\[
\dfrac{\kappa_1\,dx_1}{\sqrt{P(x_1)}}=\dfrac{\kappa_2\,dx_2}{\sqrt{P(x_2)}},
\]
where $P$ is an arbitrary quartic and $\kappa$'s are integer \cite{eul68}. In fact, the Laplace integral for the Kepler model and entries of the angular momentum for the oscillator are particular cases of the Euler integral.

Remind, that with geometric point of view Euler (and later Abel in his famous theorem) studied
the points of intersection of the hyperelliptic curve $\mu^2=P(x)$  with
any arbitrary algebraical curve whose equation in a rational form may
be written as $F(x_1,\ldots,x_n)=0$ (see the following symbolic picture for the Euler two-dimensional case) \cite{gr}.
%\begin{figure}
\begin{center}
\includegraphics{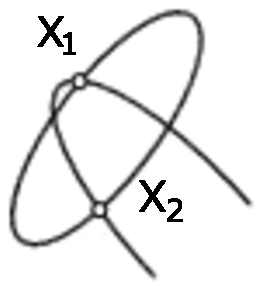}
\end{center}
%\end{figure}
\par\noindent
In the Jacobi separation of variables method these points are the points of intersection of the \textit{separated relations} $p_j^2=P(x_j)$ defined on the whole phase space with the classical  \textit{trajectory of motion} $F(x_1,\ldots,x_n)=0$ on its configurational subspace.

So,  Euler proposed algebraic construction of the classical trajectories of motion without any integration and inversion of the Abel map. This particular algebraic construction of the trajectories of motion was one of the starting points of the Jacobi investigations of the last multiplier theory and of the inversion of the Abel map.

Such as algebraic curve $F(x_1,\ldots,x_n)=0$ has many different parameterizations $x_k(t)$, there are  many different superintegrable systems associated with this curve. For instance, the overwhelming majority of the known two-dimensional superintegrable systems are related with the different parametrizations of the conic sections.
\begin{rem} For the ellipse
\[
F(x_1,x_2)=\dfrac{x_1^2}{a^2}+\dfrac{x_2^2}{b^2}-1=0,
\]
we can use well known parametrizations by trigonometric or elliptic functions
\[
x_1=a\sin t,\qquad x_2=b\cos t\qquad\mbox{or}\qquad
x_1=a\,\mathrm{sn}\,\widetilde{t},\qquad x_2=b\,\mathrm{cn}\, \widetilde{t}
\]
which may be associated with the oscillator and the Kepler problem, respectively.
\end{rem}

\subsection{Additional integrals of motion and angle variables}
A classical  superintegrable system  is an integrable
$n$-dimensional Hamiltonian system with $n+k$ single valued  and functionally independent  integrals of  motion on the whole phase space $M$. At $k=n-1$ we have maximally superintegrable systems with the closed trajectories of motion.

The Liouville classical theorem on completely integrable Hamiltonian systems implies that almost all points of the manifold $M$ are covered by a system of open toroidal domains with the action-angle coordinates $I=(I_1,\ldots,I_k)$ and $\omega=(\omega_1,\ldots,\omega_n)$:
\bq\label{aa-br}
\{I_j,I_k\}=\{\omega_i,\omega_k\}=0,\qquad \{I_i,\omega_j\}=\delta_{ij}.
\eq
The $n$ integrals of motion $H_1,\ldots,H_n$ are functions of the action variables $I_1,\ldots,I_n$ such that the corresponding Jacobian does not equal to zero
\bq\label{h-mat}
\det\mathbf J\neq 0,\qquad\mbox{where} \qquad \mathbf J_{ij}= \dfrac{\partial H_i(I_1,\ldots,I_n)}{\partial I_j}\,.
\eq
The $n-1$ additional integrals of motion
\bq\label{phi_k}
\phi_j=\sum_{k} \left(\mathbf J^{-1}\right)_{kj}\,\omega_k,\qquad
\{H_1,\phi_j\}=0,\qquad j=2,\ldots,n,
\eq
are functionally independent on $H_1(I),\ldots,H_n(I)$ and  any completely integrable system is maximally superintegrable system in a neighborhood of every regular point of $M$ \cite{ts07f}.

In generic case the  action variables $I_k$ are single-valued functions, whereas the
angle variables $\omega_k$ are multi-valued functions on $M$. However,  additional functionally independent integrals of motion $K_i$ have to be some functions on $\phi_j$, which are sums of the multivalued functions $\omega_k$.

So, for any superintegrable system we have the following question:
\par\noindent
\begin{quote}
 \textbf{How are single-valued functions obtained from the sums of multi-valued functions?}
\end{quote}
\par\noindent
In fact for all the known superintegrable systems we have only one answer:
\par\noindent
\begin{quote}
 \textbf{We have to apply addition theorem and to get one multi-valued function on the single-valued argument, which is the desired  integral of motion:}
\end{quote}
\[
\sum_{k} \left(\mathbf J^{-1}\right)_{kj}\,\omega_k=\phi_i(K_j),\qquad
\Rightarrow
\qquad
\{H_1,K_j\}=0.
\]
\par\noindent
So, the addition theorems  could help us to classify algebraically
superintegrable systems and vice versa we can get addition theorems by using known superintegrable systems.

%%%%%%%%%%%%%%%%%%%%%%%%%%%%%%%%%%%%%%%%%%%%%%%%%%%%%%%%%%%%%%%%%%%%%%%%%%%%%%%%%5
%%%%%%%%%%%%%%%%%%%%%%%%%%%%%%%%%%%%%%%%%%%%%%%%%%%%%%%%%%%%%%%%%%%%%%%%%%%%%%%%%%
\section{The  action-angle variables for the open lattices.}
\setcounter{equation}{0}
In 1975 Moser proposed construction of the action-angle variables for the open Toda lattice.
This construction may be extended on  other integrable systems related with the quadratic $r$-matrix algebras \cite{ts07f}.

For all these systems the action variables $I_j$ are zeroes of the polynomial
\bq\label{int-open}
A(\lambda) = \lambda^n +H_1\lambda^{n-1}+\cdots H_n=\prod_{j=1}^n(\lambda-I_j),\qquad \{A(\lambda),A(\mu)\}=0\,,
\eq
whose coefficients are independent integrals of motion $H_1,\ldots,H_n$ in the involution.

The angle variables $\omega_j$ are defined by the second polynomial $B(\lambda)$
having the standard Poisson bracket with the first polynomial  $A(\mu)$ \cite{ts07f}
\[
 \left\{B(\lambda),A(\mu)\right\}=\dfrac{\eta}{\lambda-\mu}\Bigl(B(\lambda)A(\mu)-B(\mu)A(\lambda)
 \Bigr),\qquad\eta\in\mathbb C,
 \]
 or
\[ \{B(\lambda),A(\mu)\}=
\frac{\scriptstyle \eta}{\scriptstyle\lambda-\mu}\Bigl( B(\lambda)A(\mu)-B(\mu)A(\lambda)\Bigr)
+\frac{\scriptstyle\eta}{\scriptstyle\lambda+\mu}\Bigl( B(\lambda)A(\mu)+B(-\mu)A(\lambda)\Bigr)\,.
\]
In this case angle variables $\omega_j$ are equal to
\bq
\omega_j=\eta^{-1}\,\ln B(I_j),\qquad j=1,\ldots,n,\qquad \eta\in\mathbb C,
\label{mos-var}
\eq
Usually  brackets between these two polynomials are either part of the Sklyanin algebra or the reflection equation algebra. It allows us to use the representation theory  in order to get single-valued function $A(\lambda)$ and $B(\lambda)$ on $M$ in initial physical variables, see \cite{ts07f}.

Using the angle variables $\omega_j$ (\ref{mos-var}) we can obtain additional algebraic integrals of motion
\[
K_j=e^{\eta\,\omega_j}=B(I_j),\qquad \{I_k,K_j\}=0,\quad k\neq j,
\]
for these maximally superintegrable systems.

Among such maximally superintegrable systems we can find the open Toda lattices associated
with the classical root systems $\mathcal A_n$, $\mathcal BC_n$, and $\mathcal D_n$ , the $XXX$ Heisenberg magnets with boundaries,
and the discrete self-trapping model with boundaries \cite{ts07f}.

\begin{rem}
The  properties of the trajectories of motion is independent on the choice of
integrals of motion in the involution  $I_1,\ldots,I_n$ or
$H_1,\ldots,H_n$. This choice  affects on the \textit{parametrization} only, i.e. on the explicit
dependence of the coordinates $q_j$ on the parameter $t$.
\end{rem}

\begin{rem}
The global action-angle variables, the corresponding Birkhoff coordinates and the KAM-theorem  for the periodic Toda lattices are discussed in \cite{hk08}. Up to now the computer experiments
indicate that periodic Toda lattice can not be superintegrable system on the whole phase space.
\end{rem}

%%%%%%%%%%%%%%%%%%%%%%%%%%%%%%%%%%%%%%%%%%%%%%%%%%%%%%%%%%%%%%%%%%%%%%%%%%%%%%%%%5
%%%%%%%%%%%%%%%%%%%%%%%%%%%%%%%%%%%%%%%%%%%%%%%%%%%%%%%%%%%%%%%%%%%%%%%%%%%%%%%%%%
\section{The action-angle variables for the St\"{a}ckel systems.}
\setcounter{equation}{0}

The system associated with the name of St\"{a}ckel \cite{st95,ts99} is a holonomic system on the phase space $M=\mathbb R^{2n}$, with the canonical variables $q=(q_1,\ldots,q_n)$ and $p=(p_1,\ldots,p_n)$:
\bq \Omega=\sum_{j=1}^n dp_j\wedge dq_j\,,\qquad
\{p_j,q_k\}=\delta_{jk}\,.\label{stw}
\eq
The nondegenerate $n\times n$ St\"{a}ckel matrix $S$, whose $j$ column depends on the coordinate $q_j$ only, defines $n$ functionally independent integrals of motion
\bq
H_k=\sum_{j=1}^n ( S^{-1})_{jk}\Bigl(p_j^2+U_j(q_j)\Bigr)\,.
\label{fint}
\eq
From this definition one immediately
gets the separated relations
\bq
p_j^2=
\sum_{k=1}^n H_k S_{kj}-U_j(q_j)\,\label{stc}
\eq
and the angle variables
\[
\omega_i=\sum_{j=1}\int\,\dfrac{S_{ij}\,\mathrm dq_j}{p_j}=
\sum_{j=1}\int\,\frac{S_{ij}\,\mathrm dq_j}{\sqrt{\sum_{k=1}^n H_k S_{kj}-U_j(q_j)  } }\,.
\]
It allows reducing  solution of the equations of motion to a problem in algebraic geometry \cite{ts99}. Namely, let us suppose that there are functions $\mu_j$ and $\lambda_j$ on the canonical separated variables
\bq\label{fr-2}
\mu_j=u_j(q_j)p_j,\qquad \lambda_j=v_j(q_j)\,,\qquad \{q_i,p_j\}=\delta_{ij},
\eq
which allows us to rewrite separated equations (\ref{stc}) as equations defining the hyperelliptic curves
\bq\label{sthc}
\mathcal C_j:\quad \mu_j^2=P_j(\lambda_j)\equiv u_j^2(\lambda_j)\left(\sum_{k=1}^n H_k S_{kj}(\lambda_j)-U_j(\lambda_j)\right),
\eq
where $P_j(\lambda_j)$ are polynomials on $\lambda_j$. In this case
the action variables $I_k=H_k$ (\ref{fint}) have the canonical Poisson brackets (\ref{aa-br}) with the  angle variables
\bq\label{w-st}
\omega_i=\sum_{j=1}^n\int_{A_j} \dfrac{S_{ij}(\lambda_j)}{\sqrt{P_j(\lambda_j)\,}}\,\mathrm d\lambda_j=\sum_{j=1}^n \vartheta_{ij}(p_j,q_j),
\eq
which are usually the sums of integrals $\vartheta_{ij}$ of the first kind Abelian differentials on the hyperelliptic curves $\mathcal C_j$ (\ref{sthc})  \cite{ts99,ts07f}.

\begin{rem}
If the separated relations are obtained from the hyperelliptic curves by some substitution (\ref{fr-2}) the Lagrangian submanifold of the corresponding St\"ackel system is a direct product of these hyperelliptic curves
\[
\mathcal F=\mathcal C_1\times \mathcal C_2\times \cdots\times \mathcal C_n.
\]
If these curves are equal $\mathcal C_j=\mathcal C$ then $\mathcal F$ may be identified with the Jacobian  $J(\mathcal C)$ of $\mathcal C$ \cite{ts99}.
\end{rem}

\subsection{Addition theorems}

In generic case the action variables (\ref{w-st}) are the sum of the multi-valued functions $\vartheta_{ij}$. However, if we are able to apply some addition theorem to the calculation of $\omega_i$ (\ref{w-st})
\bq\label{add-th}
\omega_i=\sum_{j=1}^n \vartheta_{ij}(p_j,q_j)=\Theta_i \bigl(K_i\bigr)+const,
\eq
where $\Theta_i$ is a multi-valued function on the algebraic argument $K_i(p,q)$, then one could get additional single-valued integral of motion $K_i$.

Treating $\lambda_{n+1}$ as a function of the independent variables $\lambda_1,\ldots,\lambda_n$
the addition theorem (\ref{add-th}) may be rewritten in the integral form
\[
\sum_{j=1}^n \int^{\lambda_j} \dfrac{S_{ij}(\lambda)}{\sqrt{P_j(\lambda)\,}}\mathrm d\lambda_j-
\int^{\lambda_{n+1}}\dfrac{S(\lambda)}{\sqrt{P_{n+1}(\lambda)\,}}\mathrm d\lambda_{n+1}=const
\]
or in the  differential form
\begin{equation}\label{ab-th}
\epsilon_1 \frac{S_{i1}(\lambda_1)\,\mathrm d\lambda_1}{\sqrt{P_1(\lambda_1)}}+\cdots+ \epsilon_n
\frac{S_{in}(\lambda_n)\,\mathrm d\lambda_n}{\sqrt{P_n(\lambda_n)}}-\epsilon_{n+1}
\frac{S(\lambda_{n+1})\,\mathrm d\lambda_{n+1}}{\sqrt{P_{n+1}(\lambda_{n+1})}}=0,\qquad\epsilon_i = \pm 1,
\end{equation}
where choice of the $\epsilon_i$ is related with the choice of the contours of integration.
In this case additional integrals of motion are $\lambda_{n+1}$ and $\frac{S(\lambda_{n+1})}{\sqrt{P_{n+1}(\lambda_{n+1})}}$, which have to be single-valued functions on the whole phase space.

\begin{rem}
If  we add the necessary words to the equation (\ref{ab-th}) at $P_i=P_j$, for all the $i,j=1,\ldots,n$ we obtain the famous Abel theorem, see as an example \cite{gr}. Recall, that Abel reduced equation (\ref{ab-th}) to an algebraic identity (trajectory of motion), which is a consequence of an expansion into partial fractions.
\end{rem}

Another standard form of the addition theorem for the function $f(x)$ looks like
\bq\label{add-th2}
F\Bigl(f(x),f(y),f(x+y)\Bigr)=0\,, \qquad \forall x,y\in M,
\eq
where $F$ is some function associated with $f$. For instance, we can show different forms (\ref{add-th}) and (\ref{add-th2}) for the simplest addition theorem
\bq\label{add-ln}
\ln(x)+\ln(y)=\ln\left(xy\right)\qquad \mathrm{or}\qquad e^x\e^y=\e^{x+y},
\eq
associated with the zero-genus hyperelliptic curves $\mathcal C_j$
\bq\label{fr-1}
\mathcal C_j:\qquad \mu_j^2=P_j(\lambda_j)=f_j\lambda_j^2+g_j\lambda_j+h_j,\qquad j=1,\ldots,n,
\eq
where $f_j,g_j,h_j$ are linear functions on $n$ integrals of motion $H_1,\ldots,H_n$.

Namely, if the separated relations are obtained from the zero-genus hyperelliptic curves  (\ref{fr-1})
 and $S_{ij}=\kappa_{ij}$ are constants we have
\ben
\omega_i&=&\sum_{j=1}^n
\vartheta_{ij}=\sum_{j=1}^n\int \dfrac{\kappa_{ij}}{\sqrt{f_j\lambda_j^2+g_j\lambda_j+h_j\,}}\,\mathrm d\lambda\nn\\
&=&\sum_{j=1}^n
\ln\left(\mu_j+\dfrac{2f_j\lambda_j+g_j}{2\sqrt{f_j}}\right)^{\kappa_{ij}\,f_j^{-1/2}}\nn\\
&=&\ln \prod_{j=1}^n\left(\mu_j+\dfrac{2f_j\lambda_j+g_j}{2\sqrt{f_j}}\right)^{\kappa_{ij}\,f_j^{-1/2}}=\ln K_i\,.
\nn
\en
So, the angle variables $\omega_i$ are logarithms  as for the Moser systems (\ref{mos-var}). Of course, if we want to get algebraic or polynomial integrals of motion $K_i$ we have to impose some constrains on  $\kappa_{ij}$ and $f_j$.

We have to underline that we use different hyperelliptic curves $C_j$ in this example in contrast with the Euler example.

\subsection{Trajectories of motion}

According to the standard Jacobi procedure of inversion of the Abel map the variables $q_j$ are found from the equations
\[
\sum_{j=1}^n\int \dfrac{S_{ij}(q_j)}{\sqrt{P_j(q_j,\alpha_1,\ldots,\alpha_n)\,}}\,\mathrm dq_j=\beta_i,\qquad \,i=1,\ldots,n,
\]
where $\beta_1=t$ and $H_j=\alpha_j$. From these equations one gets trajectories in \textit{parametric} form $q_j(t)$, i.e. coordinates $q_j$ are functions on the time $\beta_1=t$
and parameters $\alpha_1,\ldots,\alpha_n,\beta_2\ldots,\beta_n$.

On the other hand, for the superintegrable systems if we substitute momenta from the separated equations (\ref{stc})
\[
p_j=\sqrt{\sum_{k=1}^n \alpha_k S_{kj}(q_j)-U_j(q_j)}
\]
into additional algebraic integrals of motion $K_j$ one gets trajectories of the motion in the \textit{algebraic} form
\bq\label{cl-tr}
F_j=K_j(q_1,\ldots,q_n,\alpha_1,\ldots,\alpha_n)-k_j=0,\qquad j=1,\ldots,n-1.
\eq

\begin{rem}
There are many different parameterizations of the algebraic curves. Since,  trajectories
may be associated with the different St\"ackel systems (\ref{cl-tr}),
which are related by the canonical transformation of the time $t\to\widetilde{t}$:
\bq\label{t-change}
d\widetilde{t}=v(q)dt,\qquad v(q)=\dfrac{\det{S}}{\det\widetilde{S}}.
\eq
Here $S$ and $\widetilde{S}$ are the St\"ackel matrices for the dual systems with common trajectories \cite{ts99a,ts01}. Existence of the such dual systems is related with the fact that the Abel map is surjective and generically injective if $n=\mathrm g$ only.

Of course, such dual St\"ackel systems have so-called equivalent metrics.
\end{rem}

%%%%%%%%%%%%%%%%%%%%%%%%%%%%%%%%%%%%%%%%%%%%%%%%%%%%%%%%%%%%%%%%%%%%%%%%%%%%%%%%%%%%%%
%%%%%%%%%%%%%%%%%%%%%%%%%%%%%%%%%%%%%%%%%%%%%%%%%%%%%%%%%%%%%%%%%%%%%%%%%%%%%%%%%%%%%%
\section{Logarithmic angle variables}
\setcounter{equation}{0}

In order to classify two-dimensional superintegrable systems  associated with the addition theorem (\ref{add-ln}) we can start with a pair of the Riemann surfaces
\bq\label{dr-eq}
\mathcal C_{j}:\qquad\mu^2=P_j(\lambda)= f\lambda^2+g_j\lambda+h_j,\qquad j=1,2,
\eq
where
\[f=\alpha H_1+\beta H_2+\gamma,\quad
  g_j=\alpha^g_jH_1+\beta^g_j H_2+\gamma^g_j,\quad
  h_j=\alpha^h_jH_1+\beta^h_j H_2+\gamma^h_j,\]
and $\alpha$, $\beta$ and $\gamma$ are real or complex numbers.

We can use addition theorem (\ref{add-ln}) if we we  fix last row of the St\"ackel matrix only. Namely, substituting
\bq\label{m-rest}
S_{2j}(\lambda)=\kappa_j
\eq
into the (\ref{stc}-\ref{w-st}) one gets
\bq\label{th-eq}
\vartheta_{2j}=\int \dfrac{S_{2j}(q_j)\,\mathrm dq_j}{p_j}=
\int\dfrac{\kappa_j\,\mathrm d\lambda_j}{\sqrt{P_j}}=
\kappa_jf^{-1/2}\,\ln\left(\mu_j+\dfrac{2f\lambda_j+g_j}{2\sqrt{f}}\right)
\,,
\eq
so the angle variable
\[
\omega_2=\dfrac{1}{\sqrt{f}}\,\ln \left[\left(p_1{u_1}+ \dfrac{P_1'}{2\sqrt{f}}\right)^{\kappa_1}
\left(p_2{u_2}+\dfrac{P_2'}{2\sqrt{f}}\right)^{\kappa_2}\right],
\qquad
P^{\,\prime}_{j}=\left.\frac{d P_{j}(\lambda)}{d\lambda}\right|_{\lambda=v_j(q_j)},\]
is the multi-valued function on the desired algebraic argument
\[
K=\left(p_1{u_1}+ \dfrac{P_1'}{2\sqrt{f}}\right)^{\kappa_1}
\left(p_2{u_2}+\dfrac{P_2'}{2\sqrt{f}}\right)^{\kappa_2}.
\]
If $\kappa_{1,2}$ are positive integer, then
\bq\label{k-kappa}
K=\left(\dfrac{1}{2\sqrt{f}}\right)^{\kappa_1+\kappa_2}\left(K_{\ell}+\sqrt{f}\,K_m\right)
\eq
is the generating function of polynomial integrals of motion $K_m$ and $K_\ell$  of the $m$-th  and $m\pm 1$-th order in the momenta, respectively. As an example we have
\bq\label{k-m}
\begin{array}{ll}
K_m=2\,(p_1{u_1}\,P'_2+p_2{u_2}\,P'_1),\quad &\kappa_1=1,\, \kappa_2=1,\\
K_m=2P'_2\,\Bigl(2p_2u_2\,P'_1+p_1u_1\,P'_2\Bigr)+8p_1u_1p_2^2u_2^2f\,,
\quad&\kappa_1=1,\,\kappa_2=2,\\
K_m=2{P'_2}^2\Bigl(3p_2u_2P'_1+p_1u_1P'_2\Bigr)+8p_2^2u_2^2\Bigl(p_2u_2P'_1+3p_1u_1P'_2\Bigr)f,
\quad&\kappa_1=1,\,\kappa_2=3,
\end{array}
\eq
where $m=1,3$, $m=3,5$ ð $m=3,7$, because $P'_{1,2}$ and $f$ are linear functions on  $H_{1,2}$, which are the second order polynomials on momenta.

The corresponding expressions for the $K_\ell$ look like
\bq\label{k-ell}
\begin{array}{ll}
K_{\ell}=P'_1P'_2+4p_1p_2{u_1u_2}f,\quad &\kappa_1=1,\, \kappa_2=1,\\
K_{\ell}=P'_1{P'_2}^2+4fp_2u_2(p_2u_2P'_1+2p_1u_1P'_2)\,,
\quad&\kappa_1=1,\,\kappa_2=2,\\
K_{\ell}=P'_1{P'_2}^3+12P'_2p_2u_2(P'_2p_1u_1+P'_1p_2u_2)f+16p_1u_1p_2^3u_2^3f^2\,.
\quad&\kappa_1=1,\,\kappa_2=3.
\end{array}
\eq
Of course, we can try to get another polynomial integrals of motion using the following recurrence relations
\bq\label{rec-H}
K_{j-1}=\{K_j,H_2\},,\qquad\mathrm{and}\qquad \{H_1,K_{j}\}=0\,,\qquad j=m,m\pm1,m\pm2,\ldots.
\eq
see examples in \cite{ts08}.

Remind, that we get polynomial integrals of motion $K_m$ and $K_\ell$ for the special St\"ackel matrices only. The imposed necessary condition (\ref{m-rest}) leads to some restrictions on the substitutions  \[\lambda_j=v_j(q_j),\qquad \mu_j=u_j(q_j)p_j.\]

In fact, after these substitutions we obtain  the following expression for the St\"{a}ckel matrix
\bq\label{s-fin}
S(q)=\left(\begin{array}{cc}
\dfrac{\alpha v_1^2+\alpha^g_1 v_1+\alpha^h_1}{u_1^2} & \dfrac{\alpha v_2^2+\alpha^g_2 v_2+\alpha^h_2}{u_2^2}\\
\\
\dfrac{\beta v_1^2+\beta^g_1 v_1+\beta^h_1}{u_1^2}& \dfrac{\beta v_2^2+\beta^g_2 v_2+\beta^h_2}{u_2^2}
\end{array}\right)\,,\qquad \det S\neq 0.
\eq
So,  for a given $\kappa_{1,2}$ expressions for $\vartheta_{2j}$ (\ref{th-eq}) yield two differential equations on functions $u,v$ and parameters $\beta$:
\bq\label{uv-eq}
S_{2j}(q_j)=\dfrac{\kappa_j\,v'_j(q_j)}{u_j(q_j)}\quad \Longrightarrow\quad \kappa_j u_jv'_j=\beta v_j^2+\beta^g_jv_j+\beta^h_j,\qquad j=1,2.
\eq
There are equations on $u,v$ and parameters $\beta$ because we have to solve these equations in the fixed functional space \cite{ts08}.

\begin{prop}
If $\kappa_{j}\neq 0$ equations (\ref{uv-eq}) have the following three monomial solutions
\bq \label{uv-sol}
\begin{array}{llll}
\mathrm I\qquad&\beta=0,\quad \beta_j^h=0,\qquad& u_j=q_j,\qquad &v_j=q_j^{\frac{\beta_j^g}{\kappa_j}},
\\
\\
\mathrm{II}\qquad&\beta_j^g=0,\quad\beta_j^h=0,\qquad &u_j=1,\qquad& v_j=-\kappa_j(\beta\, q_j)^{-1},\\
\\
\mathrm{III}\qquad&\beta=0,\quad\beta_j^g=0,\qquad &u_j=1,\qquad &v_j=\kappa_j^{-1}\beta_j^h\, q_j,
\end{array}
\eq
up to canonical transformations. The fourth solution $\mathrm{(\,IV)}$ is the combination of the  first and third solutions for the different $j$'s.
\end{prop}
Below we mark all the superintegrable systems by number of the solution from this Proposition.

If we suppose that after point transformation
\bq \label{z-dr}
\begin{array}{ll}
x=z_1(q), &  y=z_2(q),\\
p_x=\mathrm w_{11}(q)p_1+\mathrm w_{12}(q)p_2,\qquad &p_y=\mathrm w_{21}(q)p_1+\mathrm w_{22}(q)p_2,\end{array}
\eq
the kinetic energy has some fixed form
\bq\label{metr-eq}
T=\sum\left(S^{-1}\right)_{1j}p_j^2=\mathrm{g}_{11}(x,y)p_x^2
+\mathrm{g}_{12}(x,y)p_xp_y+\mathrm{g}_{22}(x,y)p_y^2,
\eq
then we will get some additional restrictions on the separated relations and the St\"ackel matrix.

For instance, if we suppose that
\[T=\sum\left(S^{-1}\right)_{1j}p_j^2=p_xp_y,
\]
then  one will get the following  algebraic equations
\bq\label{alg-dr}
\mathrm w_{11}\mathrm w_{21}=\left(S^{-1}\right)_{11},\qquad
\mathrm w_{12}\mathrm w_{21}+ \mathrm w_{11}\mathrm w_{22}=0,\qquad
\mathrm w_{12}\mathrm w_{22}=\left(S^{-1}\right)_{12}
\eq
 and  the  partial differential equations
\bq\label{pde-dr}
\{x,p_x\}=\{y,p_y\}=1,\qquad \{p_x,y\}=\{p_y,x\}=\{p_x,p_y\}=0.
\eq
on  parameters $\alpha$ and  functions  $z_{1,2}(q_1,q_2)$, $\mathrm w_{kj}(q_1,q_2)$.

The remaining free parameters $\gamma,\gamma_j^h,\gamma_j^g$ determine the corresponding potential part of the Hamiltonian $V(x,y)$. In fact, since integrals $H_{1,2}$ is defined up to the trivial shifts $H_k\to H_k+c_k$, our potential $V(x,y)$ depends on three arbitrary parameters only.

\subsection{Examples}
According to  \cite{ts08} in this section we consider superintegrable systems on a complex Euclidean space $E_{2}(\mathbb{C})$  with the following Hamiltonian
\bq\label{ham-dr}
H_1=p_xp_y+V_Z(x,y)\,,
\eq
where subscript $Z=I,II,III$ or $IV$ indicates on the type of the solution (\ref{uv-sol}).
The passage from the conformal coordinate system
$(x, y)$ to another coordinate systems can be realized by using
the Beltrami partial differential equations.

\begin{rem}
According to \cite{ts99a,ts01} the St\"ackel transformations (\ref{t-change}) relate systems associated with the same solution of (\ref{uv-sol}) because the corresponding St\"ackel matrices $S_Z(q)$ are differed on the first row only.
\end{rem}

\begin{exam}
\par\noindent
Let us put $\kappa_1=1$ and $\kappa_2=1$ in (\ref{uv-eq}) and try to solve  equations (\ref{alg-dr})-(\ref{pde-dr}). Here is one system with first order additional integral $K_m$ (\ref{k-m})
\[
V_{III}^{(1)}=\gamma_1xy+\gamma_2(x+y)+\gamma_3(x-y),\nn\\
\]
and seven systems with cubic integral of motion $K_m$ (\ref{k-m})
\ben
V_{I}^{(1)}&=&\dfrac{\gamma_1  }{\sqrt{xy}}+\dfrac{\gamma_2}{(y-x)^2}+\dfrac{\gamma_3\,(y+x)}{\sqrt{xy}\,(y-x)^2}\,,
\nn\\
\quad
V_{I}^{(2)}&=&\gamma_1  \,xy+\dfrac{\gamma_2}{(y-x)^2}+\dfrac{\gamma_3} {(y+x)^2}\,,
\nn\\
V_{II}^{(1)}&=&\dfrac{\gamma_1  }{\sqrt{y(x-1)\,}}+\dfrac{\gamma_2}{\sqrt{y(x+1)\,}}
+\dfrac{\gamma_3 x}{\sqrt{x^2-1\,}}\,,
\nn\\
 V_{II}^{(2)}&=&\gamma_1  \,xy +\gamma_2 y\dfrac{2x^2+1}{\sqrt{x^2+1\,}}+\dfrac{\gamma_3 x}{\sqrt{x^2+1\,}}\,,\nn\\
V_{III}^{(2)}&=&\dfrac{\gamma_1  }{\sqrt{xy\,}}+\dfrac{\gamma_2}{\sqrt{x\,}}
+\dfrac{\gamma_3} {\sqrt{y\,}}\,,\nn\\
\quad V_{III}^{(3)}&=&\gamma_1   y^{-1/2} +\gamma_2 x y^{-1/2} +\gamma_3 x\,,\nn\\
V_{IV}&=&\dfrac{\gamma_1  }{(y+x)^2}+\gamma_2(y-x)+\dfrac{\gamma_3(3y-x)(y-3x)}3\,,
\nn
\en
Solution of the recurrence relations (\ref{rec-H})
\bq
K_{m-1}=\pm\Bigl(2p_1p_2u_1u_2+2v_1v_2f+v_1g_2+v_2g_1\Bigr)\label{int-k2}
\eq
is the additional polynomial integral of motion, which is the first order polynomial  for the system with potential  $V^{(1)}_{III}$ and the second order polynomial  in the momenta for another systems.

All these systems are listed in the Drach papers \cite{dr35}.
\end{exam}

\begin{exam}
\par\noindent
 At $\kappa_1=1$ and $\kappa_2=2$  there is one superintegrable system with cubic additional integral $K_m$ (\ref{k-m}) and quadratic integral $K_\ell$ (\ref{k-ell})
\[
V_{III}= \gamma_1(3x+y)(x+3y)+{\gamma_2}{(x+y)}+{\gamma_3}{(x-y)}\,,
\]
and seven systems with the real potentials
\[
\begin{array}{l}
V_{I}= \gamma_1(3x+y)(x+3y)+\dfrac{\gamma_2}{(x+y)^2}+\dfrac{\gamma_3}{(x-y)^2},
\nn\\
V_{II}^{(1)}= \gamma_1 xy +\dfrac{\gamma_2}{\sqrt{x^3y}}+\dfrac{\gamma_3}{x^2}\,,
\nn\\
V_{II}^{(2)}= \dfrac{\gamma_1}{\sqrt{xy}}+\dfrac{\gamma_2}{x^2} +\dfrac{\gamma_3 y}{x^3}\,,
\nn\\
V_{II}^{(3)}= \dfrac{\gamma_1}{\sqrt{xy}}+\dfrac{\gamma_2}{\sqrt{x^3y}}
+\dfrac{\gamma_3}{x^{5/4}y^{3/4}}\,,
\nn\\
V_{II}^{(4)}= \gamma_1 xy+\dfrac{\gamma_2 y}{x^3}+\dfrac{\gamma_3 y^3}{x^5}\,,
\nn\\
V_{IV}^{(1)}= \dfrac{\gamma_1}{\sqrt{xy}} +\dfrac{\gamma_2(\sqrt{x}-\sqrt{y})}{\sqrt{xy}}+\dfrac{\gamma_3}{\sqrt{xy}(\sqrt{x}+\sqrt{y})^2}\,,
\nn\\
V_{IV}^{(2)}= \gamma_1 xy+\gamma_2(x-y)+\dfrac{\gamma_3}{(x+y)^2}\,,
\nn
\end{array}
\]
for which integrals of motion $K_m$ and $K_\ell$ (\ref{k-m}-\ref{k-ell}) are  fifth and sixth order polynomials in the momenta.

Solution of the recurrence relations (\ref{rec-H})
\ben
K_{m-1}=\{K_m,H_2\}&=&4p_1p_2u_1u_2(2fv_2+g_2)
+4v_2(2fv_1+g_1)(f v_2+g_2)+(4fh_2+g_2^2)v_1\nn\\
&=& 4\mu_2(\mu_1P'_2+P'_1)-(4fh_2-g_2^2)\lambda_1-4h_2g_1\,.\nn
\en
is the additional polynomial integral of motion, which is the second order polynomial  for the system with potential  $V_{III}$ and the fourth order polynomial  in the momenta for another systems.

Solution of the next recurrence relation $K_{m-2}=\{K_{m-1},H_2\}$ is the rational function in  momenta.
\end{exam}

\begin{exam}
\par\noindent
Now we present some superintegrable St\"ackel systems at $\kappa_1=1$ and $\kappa_2=3$. Here is one system with cubic additional integral $K_m$ (\ref{k-m})
\[
V_{III}^{(1)}= \gamma_1(x+2y)(2x+y)+\gamma_2(x+2y)+ \gamma_3(2x+y)\,,\qquad
S=\left(
  \begin{smallmatrix}
    a & b \\
    1 & 1
  \end{smallmatrix}
\right),
\]
and seven systems with the real potentials
\[\begin{array}{l}
V_{I}=  \gamma_1(x+2y)(2x+y)+\dfrac{\gamma_2}{(x+y)^2}+\dfrac{\gamma_3}{(x-y)^2}\,,
\nn\\
V_{II}^{(1)}= \gamma_1 xy+\dfrac{\gamma_2}{x^{2/3}y^{4/3}}+\dfrac{\gamma_3}{x^{1/3} y^{5/3}},
\nn\\
V_{II}^{(2)}= \dfrac{\gamma_1}{\sqrt{xy}}+\dfrac{\gamma_2\sqrt{y}}{x^{5/2}}+\dfrac{\gamma_3 y^2} {x^{4}},
\nn\\
V_{II}^{(3)}= \dfrac{\gamma_1}{\sqrt{xy}}+\dfrac{\gamma_2}{x^{4/3}y^{2/3}}+\dfrac{\gamma_3} {x^{7/6}y^{5/6}},
\nn\\
V_{II}^{(4)}= \gamma_1 xy+\dfrac{\gamma_2 y^2}{x^4}+\dfrac{\gamma_3 y^5}{x^7},
\nn\\
V_{III}^{(2)}= \gamma_1(x^2-5x\sqrt{y\,}+4y)+\dfrac{\gamma_2 x}{\sqrt{y\,}}+\dfrac{\gamma_3}{\sqrt{y\,}}\,,\qquad
\nn\\
V_{IV}=\gamma_1(x+5y)(5x+y)+\gamma_2(x-y)+\dfrac{\gamma_3}{(x-y)^2}\,,
\nn
\end{array}\]
for which integrals of motion $K_m$ and $K_\ell$ (\ref{k-m}-\ref{k-ell}) are  seventh and eights order polynomials in the momenta.
\end{exam}

\subsection{Algebra of integrals of motion}

For all the superintegrable systems considered in the previous section the algebra of integrals of motion $H_{1,2}$ and $K_m,K_\ell$ (\ref{k-m}) is the  polynomial algebra in terms of the coefficients of the hyperelliptic curves
\bq
\{H_2,K_m\}=2K_\ell\,,\qquad \{H_2,K_\ell\}=2fK_m\,,\qquad
\{K_m,K_\ell\}=\pm \Phi_Z\,,\nn
\eq
where polynomial $F_Z$ depends on the type of solution (\ref{uv-sol}) only:
\ben
\Phi_{I}&=&2(4fh_2-g_2^2)^{\kappa_2-\kappa_1}\Bigl(4f(\kappa_1^2h_2g_1+\kappa_2^2h_1g_2)-g_1g_2(\kappa_2^2g_1+\kappa_1^2g_2)\Bigr)\,,
\nn\\
\Phi_{II}&=&4(4fh_2-g_2^2)^{\kappa_2-\kappa_1}\Bigl(4f(\kappa_2+\kappa_1)h_2h_1-\kappa_1h_1g_2^2-\kappa_2h_2g_1^2\Bigr)\mp K_m^\nu\,,
\nn\\
\Phi_{III}&=&4(4fh_2-g_2^2)^{\kappa_2-\kappa_1}\Bigl(4f(\kappa_1h_2+\kappa_2h_1)-\kappa_1g_2^2-\kappa_2g_1^2\Bigr)f\,,
\label{alg-comm}\\
\Phi_{IV}&=&2(4fh_2-g_2^2)^{\kappa_2-\kappa_1}\Bigl(4f(2\kappa_2fh_1-\kappa_1h_2g_1)-2\kappa_2fg_1^2+\kappa_1g_1g_2^2\Bigr).
\nn
\en
Here $\nu=1$ for $\kappa_1=\kappa_2$ and $\nu=2$ for $\kappa_1\neq\kappa_2$. Choice of the sign $+$ or $-$ depends on $\kappa$'s too.

The St\"ackel transformations (\ref{t-change}) relate systems associated with  one type of the solutions (\ref{uv-sol}), whereas algebra of integrals of motion is invariant with respect to such transformations.

%%%%%%%%%%%%%%%%%%%%%%%%%%%%%%%%%%%%%%%%%%%%%%%%%%%%%%%%%%%%%%%%%%%%%%%%%%%%%%%%%%%%%%
%%%%%%%%%%%%%%%%%%%%%%%%%%%%%%%%%%%%%%%%%%%%%%%%%%%%%%%%%%%%%%%%%%%%%%%%%%%%%%%%%%%%%%
\section{Elliptic angle variables}
\setcounter{equation}{0}

The first demonstration of the existence of an addition
theorem for elliptic functions is due to Euler \cite{eul68}, who showed that the differential relation
\bq\label{eul-th}
\dfrac{\mathrm{\kappa_1 dx}}{\sqrt{X}}+\dfrac{\mathrm{\kappa_2 dy}}{\sqrt{Y}}=0
\eq
connecting the most general quartic function of a variable $x$
\bq\label{X-quart}
X = a\mathrm x^4 + 4b\mathrm x^3 + 6c\mathrm x^2 +4d\mathrm x + e
\eq
and the same function $Y$  of another variable $\mathrm y$, leads to an algebraical
relation between $\mathrm x$ and $\mathrm y$, $X$ and $Y$ at integer $\kappa_{1,2}$.

As an example at $\kappa_{1,2}=1$ one gets
\[\left(\dfrac{\sqrt{X}-\sqrt{Y}}{\mathrm x-\mathrm y}\right)^2=a(\mathrm x+\mathrm y)^2+4b(\mathrm x+\mathrm y)+C
\]
where $C$ is the arbitrary constant of integration. This algebraic relation when rationalized leads to a symmetrical  biquadratic form of $\mathrm x$ and $\mathrm y$
\bq\label{bi-quad}
F(\mathrm x,\mathrm y)=a \mathrm x^2\mathrm y^2+2b\mathrm x \mathrm y(\mathrm x+\mathrm y)+c(\mathrm x^2+ 4\mathrm x\mathrm y +\mathrm y^2)+2d(\mathrm x+\mathrm y)+e=0,
\eq
which defines the conic section on the plane $(\mathrm x,\mathrm y)$, which then will be classical trajectory of motion in the configurational space.

According to \cite{gr} we could replace the constant $C$ in the Euler integral relation by $4c + 4s$, where   \bq\label{s-small}
s=\dfrac{F(\mathrm x,\mathrm y)-\sqrt{X}\sqrt{Y}}{2(\mathrm x-\mathrm y)^2}=
\dfrac{1}{4}\left(\dfrac{\sqrt{X}-\sqrt{Y}}{\mathrm x-\mathrm y}\right)^2- \dfrac{ (\mathrm x+\mathrm y)^2}4-b(\mathrm x+\mathrm y)-c,
\eq
is the famous algebraic Euler integral.  Treating $s$ as a function of the independent variables $\mathrm x$ and $\mathrm y$, one gets the following addition theorem
 \bq\label{ell-add}
\dfrac{\mathrm{dx}}{\sqrt{X}}+\dfrac{\mathrm{dy}}{\sqrt{Y}}+\dfrac{\mathrm{d} s}{\sqrt{S}}=0
\eq
Of course, the Euler addition theorem is a very special case of the  Abel theorem
 \cite{gr}. As an example, the similar explicit addition formulas for genus two hyperelliptic curves may be found in \cite{park32}.

The polynomial $S$ in (\ref{ell-add}) may be defined in the algebraic form
\bq\label{s-big}
\sqrt{S}=\frac{\Bigl(Y_1 \mathrm x+Y_2)\sqrt{X}-\Bigl(X_1\mathrm y+X_2\Bigr)\sqrt{Y}}{(\mathrm x-\mathrm y)^3},\quad
\eq
where
\[X_1=(a\mathrm x^3+3b\mathrm x^2+3c\mathrm x+d), \qquad X_2=b\mathrm x^3+3c\mathrm x^2+3d\mathrm x+e\] and similar to $Y_{1,2}$, or in the Weierstrass form
\bq\label{s-w}
S=4s^3-g_2s-g_3
\eq
where
\bq\label{g-X}
g_2= ae-4bd+3c^2,\qquad g_3= ace+2bcd-ad^2-eb^2-c^3, \eq
are the quadrivariant and cubicvariant of the quartic $X$ \cite{bp}, respectively.

For $\kappa_1=1$ and $\kappa_2=2$ we present Euler's equation of the trajectory of motion at $c=d=0$  only:
\[
F(\mathrm x,\mathrm y)=(e-a\mathrm y^4)\mathrm x+2\mathrm y\sqrt{e(a\mathrm y^4+6c\mathrm y^2+e)}\,,
\]
see page 355 in  \cite{eul68}.

\subsection{Classification of the Euler superintegrable systems}
According to  \cite{ts09}, in order to use the Euler addition theorem (\ref{ell-add})  for construction of the superintegrable St\"ackel systems we have to start with the genus one hyperelliptic curve
\[\mu^2=P(\lambda), \qquad\mbox{where}\qquad P(\mathrm x)=X,\]
and a pair of arbitrary substitutions
\[\lambda_j=v_j(q_j) \qquad \mu_j=u_j(q_j)p_j,\qquad j=1,2,\]
where $p$ and $q$ are canonical variables $\{p_j,q_i\}=\delta_{ij}$.

This hyperelliptic curve and substitutions give us a pair of the {separated relations}
\bq\label{sep-rel-l}
p_j^2\, u_j^2(q_j)=av_j(q_j)^4 + 4bv_j(q_j)^3 + 6c v_j(q_j)^2 +4dv_j(q_j) + e,\qquad j=1,2,
\eq
where coefficients $a,b,c,d$ and $e$ of the quartic $X$ (\ref{X-quart}) are linear functions of integrals of motion $H_{1,2}$:
\[a=\alpha_1H_1+\alpha_2H_2+\alpha,\qquad b=\beta_1H_1+\beta_2H_2+\beta,\quad\ldots,\quad
e=\epsilon_1H_1+\epsilon_2H_2+\epsilon.
\]
As above the separated relations (\ref{sep-rel-l}) coincide with the  Jacobi relations for the uniform St\"ackel systems \cite{st95,ts99,ts99a}
\bq
p_j=\sqrt{
\sum_{k=1}^n H_k \mathbf S_{kj}-U_j(q_j)\,}
\eq
where $\mathbf S$ is the so-called St\"ackel matrix and $U_j$ is the St\"ackel potential:
\[\begin{array}{l}
{\mathbf S}_{ij}=u_j^{-2}(\alpha_i v_j^4+4\beta_i v_1^3+6\gamma_i v_j^2+4\delta_i v_j+\epsilon_i),\\
\\
U_j= u_j^{-2}(\alpha v_j^4+4\beta v_1^3+6\gamma v_j^2+4\delta v_j+\epsilon),
\end{array}\qquad i,j=1,2.
\]
Solving these separated relations (\ref{sep-rel-l}-\ref{stc}) with respect to $H_{1,2}$ one gets pair of the St\"ackel integrals of motion in the involution
\bq
H_k=\sum_{j=1}^2 ( S^{-1})_{jk}\Bigl(p_j^2+U_j(q_j)\Bigr)\,,\qquad k=1,2,
\eq
and  {angle variables}
\bq\label{ang-v}
\omega_i=\dfrac12\sum_{j=1}^2\int\,\dfrac{\mathbf S_{ij}\,\mathrm dq_j}{p_j}=\dfrac12
\sum_{j=1}^2\int\,\frac{\mathbf S_{ij}\,\mathrm dq_j}{\sqrt{\sum_{k=1}^n H_k S_{kj}-U_j(q_j)  } }\,
\eq
canonically conjugated with the {action variables} $H_{1,2}$.

In generic case the action variables (\ref{ang-v}) are the sum of the multi-valued functions. However, if we are able to apply some addition theorem to the calculation of $\omega_2$
\[
\omega_2=\dfrac12\int^{v_1(q_1)} \dfrac{\mathbf S_{21}(\lambda)\mathrm d \lambda}{\sqrt{P(\lambda)}}+\dfrac12
\int^{v_2(q_2)} \dfrac{{\mathbf S}_{22}(\lambda)\mathrm d \lambda}{\sqrt{P(\lambda)}}=\dfrac12\int^{s} \dfrac{\mathrm d s}{\sqrt{S}},
\]
then one could get additional single-valued integrals of motion $s$ and $S$:
\[\{H_1,\omega_2\}=0\qquad \Rightarrow\qquad \{H_1,s\}=\{H_1,S\}=0\,.\]
Of course, opportunity to apply some addition theorem to computation of the angle variable $\omega_2$ and of the additional single-valued integrals of motion $s,S$ leads to some restrictions on our quartic $P(\lambda)$ and substitutions $\lambda_j=v_j(q_j), \quad \mu_j=u_j(q_j)p_j$, see \cite{ts09}.

For instance, if we want to use the Euler addition theorem (\ref{ell-add})  we have to put
\[
\mathbf S_{21}(\lambda)=1,\qquad\mathbf S_{22}(\lambda)=\pm 1.
\]
Such as
\[
\mathbf S_{ij}(q)=\dfrac{v_j'(q_j)}{u_j(q_j)}\mathbf S_{ij}(\lambda)\,
\]
these restrictions are equivalent to the following equations
\bq\label{eq-uv2}
\kappa_j u_jv'_j=\alpha_2 v_j^4+4\beta_2v_j^3+6\gamma_2 v_j^2+4\delta_2v_j+\epsilon_2\,, \qquad \kappa_1=1,\quad\kappa_2=\pm 1,
\eq
on functions $u(q),v(q)$ and coefficients $\alpha_2,\beta_2,\ldots,\epsilon_2$ of the quartic, because we have to solve these equations in some fixed functional space \cite{ts08}.

It's easy to prove that there are five monomial solutions
\bq\label{uv-sol2}
\begin{array}{llll}
\mathrm{I}\qquad&u_j=1,\qquad&v_j=q_j,\qquad &\epsilon_2\neq 0\\
\\
\mathrm{II}\qquad&u_j=q_j,\qquad &v_j=q_j^{4\delta_2\kappa_j^{-1}},\qquad &\delta_2\neq 0,\\
\\
\mathrm{III}\qquad&u_j=1,\qquad& v_j=q_j^{-1},\qquad &\gamma_2\neq 0 \\
\\
\mathrm{IV}\qquad&u_j=q_j^{-1},\qquad &v_j=q_j^{-1},\qquad &\beta_2\neq 0,\\
\\
\mathrm{V}\qquad& u_j=q_j^{-2},\qquad &v_j=q_j^{-1},\qquad  &\alpha_2\neq 0
\end{array}
\eq
up to canonical transformations of the separated variables $(p_j,q_j)$ and transformations of integrals of motion $H_j\to \sigma H_j+\rho$. Here notations $\alpha_2\neq 0$,  $\beta_2\neq 0,\ldots$ mean that other parameters are equal to zero.

As above, if we suppose that after some point transformation (\ref{z-dr})
kinetic part of the Hamilton function  has a special form (\ref{metr-eq}), then one gets some additional
restrictions (\ref{alg-dr}-\ref{pde-dr})) on the coefficients of the quartic $P(\lambda)$.

The remaining free parameters $\alpha,\ldots,\epsilon$ determine potential part of the Hamiltonian $V(x,y)$. In fact, since integrals $H_{1,2}$ is defined up to the trivial shifts $H_k\to H_k+\rho_k$, our potential $V(x,y)$ depends on three arbitrary parameters only.

Summing up, we have proved that classification of the Euler superintegrable systems on the plain is equivalent to solution of the equations (\ref{eq-uv2},\ref{alg-dr},\ref{pde-dr}). For all the possible solutions  additional quadratic in momenta integral of motion looks like
\[
K_2=s+c=\dfrac{F(v_1,v_2)-p_1u_1p_2u_2}{2(v_1-v_2)^2}+c,
\]
see  (\ref{s-small}) and additional cubic in momenta integral of motion is equal to \[K_3=\sqrt{S}\equiv\sqrt{4s^3-g_2s-g_3}\]
see (\ref{s-big}) and (\ref{s-w}).

\begin{rem}
Of course, additional integrals of motion for the $n$-dimensional superintegrable systems are related with the Abel theorem as well. As an example we present the quadratic in momenta Richelot integral \cite{rich42}
\[
s=\lambda_1^2\ldots \lambda_n^2\left(
\sum_{j=1}^n \dfrac{\mu_j}{\lambda_j^2\,\mathrm F'(\lambda_j)}\right)^2
-a_1\left(\dfrac1{\lambda_1}+\cdots+\dfrac{1}{\lambda_n}\right)
-a_0\left(\dfrac1{\lambda_1}+\cdots+\dfrac{1}{\lambda_n}\right)^2
\]
for the  uniform St\"ackel systems linearized on the Jacobian of the following hyperelliptic curve
\[
\mathrm  \mu^2=a_{2n}\lambda^{2n}+\cdots+a_1\lambda+a_0,
\]
such that the St\"ackel matrix is a lowest block of the corresponding  Brill-Noether matrix \cite{ts99,ts99a}.
Here $\mathrm F(\lambda)=\prod(\lambda-\lambda_j)$ and $\lambda_j=v_j(q_j), \quad \mu_j=u_j(q_j)p_j$ are the suitable substitutions.
\end{rem}

\subsection{Examples}
Let us find all the Euler superintegrable systems  on a complex Euclidean space $E_{2}(\mathbb{C})$
\bq\label{ham-con}
H_1=p_xp_y+V(x,y)
\eq
with the real potentials $V$. Solving equations (\ref{eq-uv2},\ref{alg-dr},\ref{pde-dr})
one gets the following five superintegrable potentials
\ben
V_1&=&{\alpha}(x+y)+{\beta}(y+3x)x^{-1/2}+\gamma x^{-1/2}\,,\nn\\
\nn\\
V_2&=&\alpha y(x+y^2)+\beta(x+3y^2)+\gamma y\,,\nn\\
\nn\\
V_3&=&\alpha(x+3y)(3x+y)+\beta(x+y)+\dfrac{\gamma}{(x-y)^2}\nn\\
\nn\\
V_4&=&\alpha xy^{-3}-\beta y^{-2}-\gamma xy,\nn\\
\nn\\
V_5&=&\dfrac{\alpha}{x^2}+\dfrac{\beta}{x^{3/2}\sqrt{y-1}}+\dfrac{\gamma}{x^{1/2}\sqrt{y-1}}.
\en
Recall, that implicitly all these systems have been found by Euler in 1761 \cite{eul68}.  Potentials $V_1$ and $V_3$ in explicit form have been found by Drach  \cite{dr35}, the ($\ell$) and ($g$) cases,  whereas potential $V_2$, $V_4$ and $V_5$  may be found  in \cite{kkp01}, as the $E_{10}$, $E_8$ and $E_{17}$ cases, respectively.

The first and fifth solutions (\ref{uv-sol2}) of the equations (\ref{eq-uv2}) are related with potentials $V_{1,2}$. The second and fourth solutions give us potential $V_3$, whereas third solution yields potentials $V_{4,5}$. For brevity we present the main stages of calculations only, all the details may be found in \cite{ts09}.

\begin{case}
If  we take first solution from (\ref{uv-sol2})
\[
u_j=\pm1,\quad v_j=\pm q_j,\qquad \Rightarrow\qquad p_1^2=P(q_1),\quad (-p_2)^2=P(-q_2)\]
and quartic
\[
P(\lambda)= -\dfrac{\alpha}{2} \lambda^4+2\beta \lambda^3+H_1\lambda^2+2\gamma \lambda +H_2,
\]
then  after the following change of variables
\bq\label{tr-xy1}
x=\dfrac{(q_1-q_2)^2}4,\quad p_x={p_1-p_2}{q_1-q_2},\quad y=\dfrac{(q_1+q_2)^2}2,\quad p_y=\dfrac{p_1+p_2}{q_1+q2}\,,
\eq
we obtain the St\"ackel integrals
\ben
H_1&=&p_xp_y+\alpha(x+y)+\dfrac{\beta(3x+y)}{\sqrt{x}}+\dfrac{\gamma}{\sqrt{x}}\,,\nn\\
H_2&=&(p_x-p_y)(p_xx-p_yy)-\dfrac{\alpha (x-y)^2}2-\dfrac{\beta(x-y)^2}{\sqrt{x}}+\dfrac{\gamma(x-y)}{\sqrt{x}},\nn
\en
 the Euler integrals of motion
\ben
K_2&=&s+c=\dfrac{p_y^2}{4}+\dfrac{\alpha x}2+\beta\sqrt{x},\nn\\
K_3&=&\sqrt{S}=\dfrac{p_y^2(p_x-p_y)}{4}+\dfrac{\alpha(p_x-p_y)x}2+
\dfrac{\beta(2p_xx-3p_yx+p_yy)}{4\sqrt{x}}+\dfrac{\gamma py}{4\sqrt{x}}.
\en
The same system may be obtained by using fifth substitution from (\ref{uv-sol2}).
\end{case}

\begin{case}
Using the same first solution (\ref{uv-sol2}) and another quartic
\[P(\lambda)=-\dfrac{\alpha}4\lambda^4-\beta\lambda^3-\dfrac{\gamma}2\lambda^2+H_1\lambda+H_2\]
 after the following change of coordinates
\[
x =\dfrac{(q_1+q_2)^2}4,\quad p_x =\dfrac{p_1+p_2}{q_1+q_2},\quad   y = \dfrac{q_1-q_2}2,\quad p_y = p_1-p_2,
\]
we  can get  superintegrable St\"ackel system
\ben
H_1&=&p_xp_y+\alpha y(x+y^2)+\beta(x+3y^2)+\gamma y,\nn\\
\nn\\
H_2&=&\dfrac{p_y^2}4+p_x^2x-yp_yp_x+\dfrac{\alpha(3y^2+x)(x-y^2)}4+2\beta y(x-y^2)+\dfrac{\gamma}2(x-y^2),\nn
\en
with the quadratic Euler integral
\[
K_2=s+c=\dfrac{p_x^2}4+\dfrac{\alpha y^2}4+\dfrac{\beta y}2
\]
and with the cubic integral of motion
\[
K_3=\sqrt{S}=\dfrac{p_x^3}4+\dfrac{\alpha\Bigl(3p_x(x+y^2)-p_yy\Bigr)}{8}+\dfrac{\beta(6p_xy-py)}8+\dfrac{\gamma p_x}8.
\]
The same system may be obtained by using fifth substitution from (\ref{uv-sol2}) as well.
\end{case}

\begin{case}
If  we take the second solution from (\ref{uv-sol2})
\[
u_j=q_j,\quad v_j=q_j^2,\qquad\Rightarrow\qquad p_1^2q_1^2=P(q_1^{2}),\quad p_2^2q_2^2=P(q_2^{2})\]
and quartic
\[
P(\lambda)=-\alpha \lambda^4-\dfrac{\beta}2 \lambda^3+H_1\lambda^2+H_2\lambda+\gamma,
\]
after canonical transformations  (\ref{tr-xy1}) we will get the following superintegrable St\"ackel system
\[\begin{array}{l}
H_1=p_xp_y+\alpha(x+3y)(3x+y)+\beta(x+y)+\dfrac{\gamma}{(x-y)^2},\\
\\
H_2=(p_x-p_y)(p_xx-p_yy)-2\alpha(x+y)(x-y)^2-\dfrac{\beta(x-y)^2}2-\dfrac{2\gamma(x+y)}{(x-y)^2}.
\end{array}
\]
Using the Euler addition theorem (\ref{ell-add}) we can get the Euler integral
\[
K_2=s+c=\dfrac{(p_x+p_y)^2}{16}+\alpha(x+y)^2+\dfrac{\beta}{4}(x+y)
\]
and the cubic in momenta integral of motion
\ben
K_3=\sqrt{S}&=&
\dfrac{(p_x-p_y)^2(p_x+p_y)}{32}
-\dfrac{\alpha(x-y)\Bigl(p_x(5x+3y)-(3x+5y)p_y\Bigr)}{8}\nn\\
&-&\dfrac{\beta(p_x-p_y)(x-y)}{16}-\dfrac{\gamma(p_x+p_y)}{8(x-y)^2}\,.\nn
\en
The same system may be obtained by using fourth substitution from (\ref{uv-sol2}).
\end{case}

\begin{case}
If  we take the third solution from (\ref{uv-sol2})
\[
u_j=\pm 1,\quad v_j=\pm q_j^{-1},\qquad\Rightarrow\qquad p_1^2=X(q_1^{-1}),\quad p_2^2=Y(-q_2^{-1})\]
and quartic
\[ P(\lambda)=\alpha\lambda^4-\beta\lambda^3+H_2\lambda^2+H_1\lambda-\gamma,
\]
then after canonical transformation
\[
x = \sqrt{q_1q_2},\quad
p_x =\dfrac{p_1q_1+q_2p_2}{\sqrt{q_1q_2}},\quad
y= \dfrac{q_1-q_2}{\sqrt{q_1q_2}},\quad
p_y = \dfrac{\sqrt{q_1q_2}(p_1q_1-q_2p_2)}{q1+q2}
\]
we will get the following superintegrable St\"ackel system
\[
H_1=p_xp_y+\dfrac{\alpha y}{x^3}+\dfrac{\beta}{x^2}+\gamma xy,\qquad
H_2=p_y^2+\dfrac{(p_xx-yp_y)^2}{4}-\dfrac{\alpha(y^2+1)}{x^2}-\dfrac{\beta y}{x}+\gamma x^2.
\]
Using the Euler addition theorem (\ref{ell-add}) one gets additional quadratic in momenta Euler integral
\[
K_2=s+C=\dfrac{(p_xx-yp_y)^2}{16}-\dfrac{\alpha y^2}{4x^2}-\dfrac{\beta y}{4x}
\]
and the cubic integral of motion
\[
K_3=\sqrt{S}=\dfrac{p_y^2(p_xx-yp_y)}{8}+\dfrac{\alpha(3yp_y+xp_x)}{8x^2}+\dfrac{\beta p_y}{4x}+\dfrac{\gamma(p_xx-yp_y)x^2}{8}.
\]

\end{case}
\begin{case}
Using the same third solution (\ref{uv-sol2}) and another quartic
\[P(\lambda)=
4\alpha \lambda^4+4\beta \lambda^3+H_2 \lambda^2+2\gamma \lambda+H_1
\]
after the following change of variables
\[
x =\dfrac{q_1q_2}2,\quad
p_x =\dfrac{p_1q_1+q_2p_2}{q_1q_2},\quad
y =\dfrac{q_1^2+q_2^2}{2q_1q_2},\quad
 p_y =\dfrac{q_1q_2(p_1q_1-q_2p_2)}{q_1^2-q_2^2},
\]
we can get more complicated superintegrable system with the St\"ackel integrals of motion
\ben
H_1&=&p_xp_y+\dfrac{\alpha}{x^2}+\dfrac{\beta}{x^{3/2}\sqrt{y-1}}+\dfrac{\gamma}{x^{1/2}\sqrt{y-1}},\nn\\
H_2&=&(xp_x-p_y-p_yy)(xp_x+p_y-p_yy)-\dfrac{4\alpha y}{x}-\dfrac{2\beta(2y-1)}{x^{1/2}\sqrt{y-1}}-\dfrac{2\gamma x^{1/2}}{\sqrt{y-1}}.\nn
\en
The Euler integral of motion is equal to
\[
K_2=s+c=\dfrac{(p_xx+p_y-p_yy)^2}{4}-\dfrac{\alpha(y-1)}{x}-\dfrac{\beta\sqrt{y-1}}{\sqrt{x}},
\]
whereas the cubic in momenta integral reads as
\[
K_3=\sqrt{S}=-\dfrac{p_y(xp_x+p_y-p_yy)^2}2+\dfrac{2\alpha(y-1)p_y}{x}
-\dfrac{\beta(p_xx-3p_yy+3py)}{2x^{1/2}\sqrt{y-1}}
-\dfrac{\gamma x^{1/2}(p_xx+p_y-p_yy)}{2\sqrt{y-1}}.
\]
\end{case}

This list of examples may be easily broadened because Euler proposed construction of the algebraic integrals of motion for the equations (\ref{eul-th})
 with any integer $\kappa$'s. We discuss the Euler construction of the algebraic integrals of motion for superintegrable systems at $\kappa_{1,2}=\pm1$ only. Of course, the most interesting case  is the case with the different $\kappa_{1,2}$ in  (\ref{eul-th}) for which the corresponding additional integrals of motion will be higher order polynomials in momenta as for logarithmic angle variables.

\subsection{The quadratic integrals of motion}
It is easy to prove that the algebra of integrals of motion $H_{1,2}$ and $K_2$ is the quadratic Poisson algebra because
\[\{H_2,K_2\}=\sigma K_3=\sigma \sqrt{S(s)},\quad\mbox{where}\quad
\left\{\begin{array}{ll}
\sigma=2,\quad& V_1,\\
\sigma=-2,\quad& V_2,V_4,V_5,\\
\sigma=4,\quad& V_3,
\end{array}
\right.
\]
and
\ben
\{H_2\{H_2,K_2\}\}=\{H_2,\sigma\sqrt{S}\}=\dfrac{\sigma^2}{2}\,S'=\dfrac{\sigma^2}{2}
\Bigl(12s^2-g_2\Bigr)=\Phi(H_1,H_2,K_2),
\en
where  $\Phi(H_1,H_2,K_2)$ is the second order polynomial such as
$s=K_2-c$ and $g_2$ is quadrivariant of the quartic (\ref{g-X}).
Another details on the quadratic Poisson algebras of integrals of motion may be found in \cite{cd06}.

The search of the two dimensional manifolds whose the geodesics are curves which possess two additional quadratic integrals of motion was initiated by  Darboux \cite{darb}, who found
five classes of the metrics. These metrics or "formes essentielles"
are tabulated in "Tableau" by Koenigs \cite{koe} and  in \cite{kkm07}.

The superintegrable Darboux-Koenigs systems have a generic conformal Hamiltonian
\[
H_1=\dfrac{p_\xi p_\eta}{\mathrm{g}(\xi,\eta)}+V(\xi,\eta),
\]
where the Darboux-Koenigs metric $\mathrm g$ is a metric on the Liouville  surface \cite{darb,cd06} if
\[
\mathrm g(\xi,\eta)=F(\xi+\eta)+G(\xi-\eta),\qquad\mbox{and}\qquad
K_2={p_\xi}^2+{p_\eta}^2-2p_\xi p_\eta\dfrac{\beta(\xi,\eta)}{\mathrm g(\xi,\eta)}+Q(\xi,\eta)
\]
or metric $\mathrm g$ is a metric on the Lie surface if
\[
\mathrm g(\xi,\eta)=\xi\,F(\eta)+G(\eta),\qquad\mbox{and}\qquad
K_2={p_\xi}^2-2p_\xi p_\eta\dfrac{\beta(\xi,\eta)}{\mathrm g(\xi,\eta)}+Q(\xi,\eta).
\]
Superintegrable systems associated with the Liouville surfaces are separable in the two different orthogonal  systems of coordinates. It means that two pairs of integrals of motion $(H_1,H_2)$ and $(H_1,K_2)$ take on the St\"ackel form (\ref{fint}) after some different point transformations (\ref{z-dr}).

For the superintegrable systems associated with the
Lie surfaces only one pair of integrals $(H_1,H_2)$ may be reduced to the St\"ackel form (\ref{fint}), whereas second pair of integrals $(H_1,K_2)$ doesn't separable in the class of the point transformations (\ref{z-dr}).

It's easy to prove that two systems with potentials $V_1$ and $V_2$ are defined on the Lie surfaces. The remaining systems with potentials $V_3,V_4$ and $V_5$ are defined on the Liouville surfaces. The second separated variables $\widetilde{q}_{1,2} $ for integrals of motion $H_1,K_2$ may be found by using the software proposed in  \cite{ts05}:
\[
\begin{array}{lll}
x=\dfrac{\widetilde{q}_1^2-\widetilde{q}_2^2}{4},\quad &y=\dfrac{\widetilde{q}_1^2+\widetilde{q}_2^2}{4},
\qquad &\mbox{for}~V_3,\\
\\
x=\widetilde{q}_2\,\widetilde{q\,}_1^{-2},\quad &y=\widetilde{q}_2\,\widetilde{q\,}_1^{2},\qquad &\mbox{for}~V_4,V_5.
\end{array}
\]
It is easy to prove that the corresponding separated relations $\widetilde{p}_j^2=P(\widetilde{q}_j)$  define two different zero-genus hyperelliptic curves and lead to logarithmic angle variables \cite{ts08}. Thus, for these three Euler  superintegrable systems there are two different addition theorems:  addition theorem for elliptic functions (\ref{ell-add}) and addition theorem for logarithms $\ln \mathrm x+\ln \mathrm y=\ln \mathrm{xy}$.  So, multiseparability of the superintegrable systems may be associated with occurrence of the different addition theorems for a given superintegrable hamiltonian.

According to \cite{imm00}, the St\"ackel integrals of motion are in the bi-involution with respect to a pair of compatible Poisson brackets
\[
\{H_i,H_j\}=\{H_i,H_j\}'=0,
\]
where the second Poisson tensor $P'=NP$ is obtained by the following recursion operator $N$:
\[
N\,\dfrac{\partial }{\partial q^k}=\sum_{i=1}^n L_k^i
\dfrac{\partial }{\partial  q^i}+\sum_{ij} p_j\left(\dfrac{\partial
L^j_i}{\partial q^k}-\dfrac{\partial L^j_k}{\partial
q^i}\right)\dfrac{\partial }{\partial  p_i},\qquad
N\,\dfrac{\partial }{\partial  p_k}=\sum_{i=1}^n L^k_i
\dfrac{\partial }{\partial p_i}.
\]
Here $L$ is so-called Benenti tensor \cite{ben97}.
There is symbolic software \cite{ts05}, which allows us to obtain this tensor
starting with the Hamiltonian $H_1$ only.

So, it is easy to prove that any superintegrable system associated with the Liouville surface has two different linear in momenta Poisson bivectors $P'$ and $P''$, which are compatible with the canonical bivector $P$ and incompatible to each other. On the other hand, superintegrable system associated with the Lie surface has one linear Poisson bivector $P'$ only. It will be interesting  to understand the geometric origin of this phenomena.

Moreover,  we can find some integrable systems associated with the Lie surfaces, which doesn't separable in the class of the point transformations and which hasn't  linear second Poisson bivectors. As an example, we present the following integrable system with the quadratic integrals of motion
\[
H_1=p_xp_y+\alpha  y^{-3/2} +\beta x y^{-3/2} +\gamma x,\qquad
K_2=p_x^2-4\beta y^{1/2}+2\gamma y\,,
\]
which hasn't any other quadratic integrals. It will be interesting to classify such systems and describe  the non-linear bi-hamiltonian structures associated with the Lie surfaces.

The research was partially supported by the RFBR grant 06-01-00140.


\begin{thebibliography}{10}
\bibitem{ben97}
S. Benenti, \newblock{\em  Intrinsic characterization of the variable separation in the Hamilton-Jacobi equation}, J. Math. Phys. v.38, pp.6578-6602, 1997.

\bibitem{bp}
W.S. Burnside and A.W. Panton, \newblock{\em Theory of Equations},  Longmans, 1886.

\bibitem{gr}
A. Cayley, \newblock{\em An Elementary Treatise on Elliptic Functions},  London, 1876.\\
A.G. Greenhill,
 \newblock{\em  The applications of elliptic functions}, London,  1892.

\bibitem{darb}
G. Darboux, Le\c{c}ons sur la Th\'{e}orie G\'{e}n\'{e}rale des Surfaces, 1898.

\bibitem{cd06}
C. Daskaloyannis, K. Ypsilantis,
\newblock{\em Unified treatment and classification of superintegrable systems with integrals quadratic in momenta on a two-dimensional manifold}, J. Math. Phys., v.47, 042904,  2006.

\bibitem{park32} A.L. Dixon,
\newblock{\em On hyperelliptic f unctions of genus two}, Quarterly Journal of Mathematics,
v. 36, p. 1, 1904.\\
W.V. Parker, \newblock{\em  Addition formulas for hyperelliptic functions},
Bulletin A. M. S. v.38, p.895-901, 1932.

\bibitem{dr35}
J.~Drach. \newblock{\em Sur l'int\'{e}gration logique des \'{e}quations de la dynamique \`{a} deux variables: Forces conservatives. Int\^{e}grales cubiques. Mouvements dans le plan.}
\newblock {Comptes Rendus},   v.200, p.22-26, 1935.


\bibitem{eul68}
L. Euler, Institutiones Calculi integralis, Ac.Sc. Petropoli, 1761,
(Russian translation GITL, Moskow, 1956.)

\bibitem{hk08}
A. Henrici, T. Kappeler,
\newblock{\em Global action-angle variables
for the periodic Toda lattice}, Preprint: arXiv:0802.4032, 2008.


\bibitem{ts05}
Yu.A. Grigoryev, A.V. Tsiganov,
\newblock{\em Symbolic software for separation of variables in the Hamilton-Jacobi equation for the L-systems}, Regular and Chaotic Dynamics, v.10(4), p.413-422, 2005.

\bibitem{ts09}
Yu A Grigoryev, V. A. Khudobakhshov, A V Tsiganov,
\newblock{\em  On the Euler superintegrable systems}, to appear.

\bibitem{mw08}
I. Marquette, P. Winternitz,
\newblock{\em Superintegrable systems with third-order integrals of motion},
J. Phys. A: Math. Theor., v.41, 304031 (10pp), 2008.

\bibitem{imm00}
A. Ibort, F. Magri, G. Marmo,
 {\em Bihamiltonian structures and St\"{a}ckel separability},
 {J. Geometry and Physics}, v.33, p.210-228, 2000.

 \bibitem{kkp01} E.G. Kalnins, J.M. Kress, G.S. Pogosyan and W. Miller,
\newblock{\em Completeness
of superintegrability in two-dimensional constant-curvature spaces},
J.Phys. A: Math. Gen. v.34, p.4705, 2001.

\bibitem{kkm07}
E.G. Kalnins, J.M. Kress, W. Miller Jr,
\newblock{\em 	Nondegenerate 2D complex Euclidean superintegrable systems and algebraic varieties}, J. Phys. A: Math. Theor., 40, 3399-3411, 2007.

\bibitem{koe}
M.G. Koenigs,
\newblock{\em Sur les g\'{e}od\'{e}siques a int\'{e}grales
quadratiques},  Note II in G. Darboux, Le\c{c}ons sur la Th\'{e}orie G\'{e}n\'{e}rale des Surfaces, 1898.



\bibitem{mos75}
J. Moser,
\newblock{\em Finitely many mass points on the line under the influence
of an exponential potential - an integrable system}, Lect. Notes in Phys.,
v.38, p.467-497, 1975.

\bibitem{rich42}
F. Richelot,
\newblock{\em Ueber die Integration eines merkw\"urdigen Systems von Differentialgleichungen},
J. Reine Angew. Math. v.23, p.354-369,1842.



\bibitem{st95}
P. St\"{a}ckel  {\em Uber die Integration der
Hamilton--Jacobischen Differential Gleichung Mittelst Separation der
Variabel}, Habilitationsschrift, Halle, 1891.

\bibitem{ts99}	A.V. Tsiganov,
\newblock{\em The St\"{a}ckel systems and algebraic curves},
J. Math. Phys., v.40, p.279-298, 1999.

\bibitem{ts99a}
A.V. Tsiganov,
\newblock{\em Duality between integrable St\"{a}ckel systems},
J. Phys.A: Math. Gen., v.32, p.7965--7982, 1999.

\bibitem{ts00}
A.V. Tsiganov,	\newblock{\em On the Drach superintegrable systems},
J. Phys.A: Math. Gen., v.33, p.7407-7423, 2000.

\bibitem{ts01}
A.V. Tsiganov,	\newblock{\em The Maupertuis principle and canonical transformations of the extended phase space}, J. Nonlinear Math.Phys, v.8(1), p.157-182, 2001.



\bibitem{ts07f}
A.V. Tsiganov,
\newblock{\em On maximally  superintegrable  systems},
Regular and Chaotic Dynamics, v.13(3), p. 178-190, 2008.

\bibitem{ts08}
A.V. Tsiganov,
\newblock{\em Addition theorem and the Drach superintegrable systems},
J. Phys. A: Math. Theor., v.41(33), 335204 (16pp), 2008.








\end{thebibliography}
\end{document}